\documentclass[aps,twocolumn,superscriptaddress]{revtex4}
\usepackage{epsfig}
\usepackage{times}
\usepackage{color}
\begin{document}

\title{Optimal interdependence between networks for the evolution of cooperation}

\author{Zhen Wang}
\affiliation{Department of Physics, Hong Kong Baptist University, Kowloon Tong, Hong Kong}
\affiliation{Center for Nonlinear Studies and the Beijing-Hong Kong-Singapore Joint Center for Nonlinear and Complex Systems, Hong Kong Baptist University, Kowloon Tong, Hong Kong}

\author{Attila Szolnoki}
\affiliation{Institute of Technical Physics and Materials Science, Research Centre for Natural Sciences, Hungarian Academy of Sciences, P.O. Box 49, H-1525 Budapest, Hungary}

\author{Matja{\v z} Perc}
\email{matjaz.perc@uni-mb.si}
\affiliation{Faculty of Natural Sciences and Mathematics, University of Maribor, Koro{\v s}ka cesta 160, SI-2000 Maribor, Slovenia}

\begin{abstract}
Recent research has identified interactions between networks as crucial for the outcome of evolutionary games taking place on them. While the consensus is that interdependence does promote cooperation by means of organizational complexity and enhanced reciprocity that is out of reach on isolated networks, we here address the question just how much interdependence there should be. Intuitively, one might assume the more the better. However, we show that in fact only an intermediate density of sufficiently strong interactions between networks warrants an optimal resolution of social dilemmas. This is due to an intricate interplay between the heterogeneity that causes an asymmetric strategy flow because of the additional links between the networks, and the independent formation of cooperative patterns on each individual network. Presented results are robust to variations of the strategy updating rule, the topology of interdependent networks, and the governing social dilemma, thus suggesting a high degree of universality.
\end{abstract}

\maketitle

Network reciprocity is amongst the most well-known mechanisms that may sustain cooperation in evolutionary games that constitute a social dilemma \cite{nowak_s06}. It was discovered by Nowak and May \cite{nowak_n92b}, who observed that on structured populations cooperators can aggregate into compact clusters and so avoid being wiped out by defectors. Although the mechanism may not work equally well for all social dilemmas \cite{hauert_n04}, and recent empirical evidence based on large-scale economic experiments indicate that it may be compromised or fail altogether \cite{gracia-lazaro_srep12, gracia-lazaro_pnas12}, there is still ample interest in understanding how and why networks influence the evolution of cooperation. Recent reviews are a testament to the continued liveliness of this field of research \cite{szabo_pr07, roca_plr09, perc_bs10, perc_jrsi13}.

Following the explorations of evolutionary games on individual small-world \cite{abramson_pre01, kim_bj_pre02, masuda_pla03, tomassini_pre06, fu_epjb07}, scale-free \cite{santos_prl05, santos_pnas06, gomez-gardenes_prl07, rong_pre07, masuda_prsb07, tomassini_ijmpc07, szolnoki_pa08, assenza_pre08, santos_n08, pena_pre09, poncela_pre11, brede_epl11, tanimoto_pre12, pinheiro_pone12, simko_pone13}, coevolving \cite{ebel_pre02, zimmermann_pre04, pacheco_prl06, santos_ploscb06, fu_pa07, tanimoto_pre07}, hierarchical \cite{lee_s_prl11} and bipartite \cite{gomez-gardenes_c11} networks, the attention has recently been shifting towards interdependent networks \cite{wang_z_epl12, wang_z_srep13, gomez-gardenes_srep12, gomez-gardenes_pre12, wang_b_jsm12}. The latter have been put into the spotlight by Buldyrev et al. \cite{buldyrev_n10}, showing that even seemingly irrelevant changes in one network can have catastrophic and very much unexpected consequence in another network. Subsequently, interdependent networks have been tested for their robustness against attack and assortativity \cite{huang_xq_pre11, zhou_d_pre12, baxter_prl12, peixoto_prl12}, properties of percolation \cite{parshani_prl10, song_epl12, lau_pre12, schneider_pre13, zhao_jsm13} and diffusion \cite{gomez_prl13}, and they have indeed become a hot topic of general interest \cite{gao_jx_np12, havlin_pst12}, touching upon subjects as diverse as epidemic spreading \cite{wang_pla12}, the appearance and promotion of creativity \cite{csermely_tde13}, and voting \cite{halu_epl13}.

Previous research concerning evolutionary games on interdependent networks has revealed, for example, that biased utility functions suppress the feedback of individual success, which leads to a spontaneous separation of characteristic time scales on the two interdependent networks \cite{wang_z_epl12}. Consequently, cooperation is promoted because the aggressive invasion of defectors is more sensitive to the deceleration. Even if the utilities are not biased, cooperation can still be promoted by means of interdependent network reciprocity \cite{wang_z_srep13}, which however requires simultaneous formation of correlated cooperative clusters on both networks. It has also been shown that the coupling of the evolutionary dynamics in each of the two networks enhances the resilience of cooperation, and that this is intrinsically related to the non-trivial organization of cooperators across the interdependent layers \cite{gomez-gardenes_srep12}. Perhaps most closely related to the setup of the present work is that by Wang et al. \cite{wang_b_jsm12}, who showed that probabilistic interconnections between interdependent networks can very much promote the evolution of cooperation. In our model, however, the strategy transfer between networks is prohibited. The interdependence is thus due solely to coupling together the payoffs of select players that reside on different networks.

Here, we wish to determine how strong the interdependence between the networks really ought to be for the optimal promotion of cooperation. Since existing works unequivocally declare that interdependence works in favor of the resolution of social dilemmas, one might intuitively assume that the stronger the interdependence the better. As we will show, however, this assumption is not necessarily true. To address the problem, we consider primarily the prisoner's dilemma game on two square lattices, where a certain fraction of randomly selected players is allowed to connect with the corresponding players in the other lattice. While strategy transfers between the two networks are not allowed, the additional connections between the corresponding players do influence their utility, and thus their ability to retain and possibly spread their strategies on the home network. This introduces two new parameters, namely the fraction of players that is allowed to form links with the corresponding players in the other network $\rho$, and the strength of this links $\alpha$. Together, these two parameters determine the strength of interdependence, and also the success of resolving social dilemmas. For further details with regards to the studied evolutionary games, the applied dynamical rule, and the topology of interaction networks, we refer to the Methods section. Independent of the strategy updating rule, the topology of interdependent networks, and the governing social dilemma, we will show that cooperation is promoted best if only an intermediate fraction of players is allowed to have external links to the other network, but also that those links should be sufficiently strong. We will also reveal mechanisms that lead to the emergence of the optimal interdependence.

\section*{Results}

\begin{figure}
\centerline{\epsfig{file=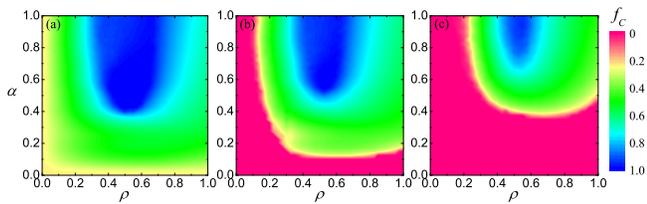,width=8.5cm}}
\caption{\label{ar_plane} Tuning in on the optimal interdependence between two square lattices for the resolution of the prisoner's dilemma. Color coded is the fraction of cooperators $f_C$ in dependence on the fraction of players that are allowed to form an external link $\rho$ and the strength of these links $\alpha$, as obtained for $b=1.03$ (a), $b=1.05$ (b) and $b=1.1$ (c). Irrespective of $b$, there exists an intermediate value of $\rho \approx 0.5$ at which cooperation is optimally promoted. But in addition to that, the value of $\alpha$ needs to be sufficiently large as well.}
\end{figure}

To begin with, we show in Fig.~\ref{ar_plane} the impact of parameters $\rho$ and $\alpha$ on the outcome of the prisoner's dilemma game. It can be observed that there exists an intermediate range of the fraction of players that are allowed to form an external link at which cooperators fare best. Irrespective of the temptation to defect $b$, values around $\rho \approx 0.5$ yield an optimal outcome of the social dilemma. Yet the coupling strength is important too. Only if the value of $\alpha$ is sufficiently large are the players able to utilize the advantage of being linked to their corresponding players in the other network. Although the level of cooperation appears to fade slightly beyond $\alpha=0.7$ if the temptation to defect is low or moderate [panels (a) and (b)], the prevailing conclusion is that the coupling strength needs to be sufficiently strong.

\begin{figure}
\centerline{\epsfig{file=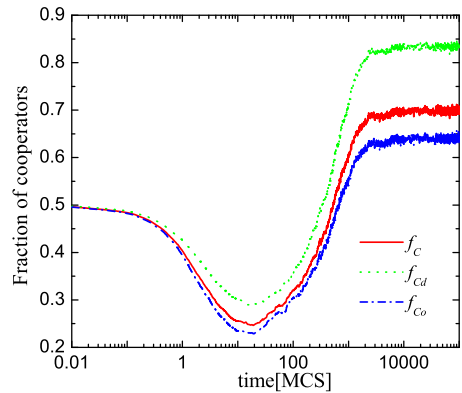,width=6cm}}
\caption{\label{evol_frac} Distinguished players who have an external link with their corresponding player in the other network are more likely to cooperate than those who are not externally linked. Depicted is the time evolution of the fraction of cooperators in the whole population ($f_C$), among the distinguished players ($f_{C_d}$), and among ordinary players, i.e., those that do not have an external link to the other network ($f_{C_o}$). It can be observed that $f_{C_d}>f_{C_o}$. Parameter values used were: $b=1.05$, $\rho=0.3$ and $\alpha=0.8$.}
\end{figure}

To clarify the mechanism that is responsible for the promotion of cooperation, we first monitor the evolution of cooperation by measuring not just the overall average cooperation level ($f_C$), but separately also the average cooperation level for players with ($f_{C_d}$) and without ($f_{C_o}$) external links to the other network in dependence on the number of full Monte Carlo steps (MCS), as defined in the Methods section. For easier reference, we will refer to individuals with external links to the other network as ``distinguished'' and to those who have no such links as ``ordinary'' players (note that the subscripts in $f_{C_d}$ and $f_{C_o}$ are chosen accordingly). Figure~\ref{evol_frac} reveals that the cooperation level amongst the distinguished players who do have external links to the other network is significantly higher than the cooperation level amongst the players who are not externally linked. Expectedly, the overall cooperation level is in-between $f_{C_d}$ and $f_{C_o}$. The identified difference between $f_{C_d}$ and $f_{C_o}$ is crucial, because it indicates that players who have the ability to collect an additional payoff from the other network are more likely to cooperate. Indeed, the natural selection of the cooperative strategy among distinguished players is higher than in the whole population, which is a consequence of an asymmetric strategy flow that emerges between the distinguished and other players. Because of generally higher payoffs, the distinguished players are followed by the others, which results in the selection of cooperation around them. In other words, distinguished players with an external link to the other network play the role of leaders in the community, similarly as was reported many times before for hubs on heterogeneous isolated (individual) networks \cite{szabo_pr07}.

\begin{figure}
\centerline{\epsfig{file=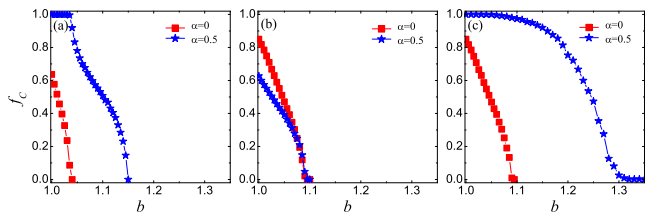,width=8.5cm}}
\caption{\label{fC_comp} Interdependence between networks favors cooperation only if the awarded additional payoffs are not counteracted by reduced teaching activity. Panel (a) depicts the fraction of cooperators $f_C$ in dependence on the temptation to defect $b$ for the basic version of the game, where all players have teaching activity $w=1$. In panel (b) the teaching activity of all distinguished players who have an external link to the other network is reduced to $w=0.05$. It can be observed that the promotion of cooperation due to the interdependence vanishes. In panel (c) the teaching activity of all ordinary players is reduced to $w=0.05$, which further strengthens the leading role of the distinguished players and leads to the strongest promotion of cooperation. All panels feature results for $\alpha=0$ and $\alpha=0.5$ at $\rho=0.5$, where $\alpha=0$ means that effectively the two networks are isolated, i.e., there is no interdependence because the links between the two networks yield no additional utility to either player.}
\end{figure}

To test this explanation further, we directly adjust the teaching activity of players, i.e., the ability to pass their strategy to a neighbor \cite{szolnoki_epl07, szolnoki_njp08}, which can be done effectively by introducing a multiplication factor $w$ to Eq.~\ref{fermi}. We consider two options. First, we depreciate all distinguished players by using $w=0.05$ while keeping $w=1$ for those who do not have an external link, and second, we reverse these values. The expectation is that the first option will nullify the advantage of interdependence between the two networks, while the second option will further amplify the positive effects on the evolution of cooperation.

Figure~\ref{fC_comp} reveals nicely how the leading role of distinguished players improves the cooperation level [panels (a) and (c)], yet only if distinguished players are not depreciated by $w=0.05$ [panel (b)]. For comparison, we show in panel (a) the results obtained with the basic model where all players have $w=1$. There the change from $\alpha=0$ to $\alpha=0.5$ (note that for $\alpha=0$ the difference between distinguished and ordinary players vanishes) introduces a noticeable increase in the critical temptation to defect where cooperators die out and an overall increase in the level of cooperation. But if $w=0.05$ is applied to distinguished players, then they become unable to lead the others despite of the fact that their utilities are higher than those of their neighbors. In the absence of hierarchy that was previously warranted by the interdependence between networks, there will form no groups with homogeneous strategies, and hence the advantage of cooperation will not be revealed. As panel (b) shows quite interestingly, $\alpha=0$ (absence of interdependence between networks) even slightly outperforms $\alpha=0.5$, because a weak hierarchy is then restored due to the unequal teaching activity of players. Note that at $\alpha=0.5$ the unequal teaching activity is counteracted by the additional payoff distinguished players receive from the other network. At $\alpha=0$ this counterbalance is lost, and the ordinary players become the leaders due to their higher value of $w$. The positive effect on the level of cooperation is, however, rather marginal.

\begin{figure}
\centerline{\epsfig{file=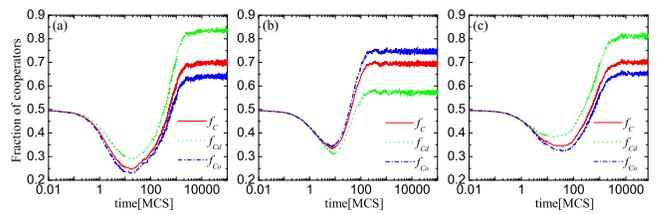,width=8.5cm}}
\caption{\label{evol_comp} Decreasing the teaching activity of distinguished players nullifies their higher propensity to cooperate, and it reduces the temptation to defect that still warrants a healthy cooperation level in the whole population. Depicted is the time evolution of the fraction of cooperators in the whole population ($f_C$), among the distinguished players ($f_{C_d}$), and among ordinary players ($f_{C_o}$). It can be observed that $f_{C_d}>f_{C_o}$ in panels (a) and (c), but not in panel (b). In the latter, the teaching activity of all distinguished players who have an external link to the other network is reduced to $w=0.05$. In panel (a) all players have $w=1$, while in panel (c) the teaching activity of all ordinary players is reduced to $w=0.05$, which further amplifies the $f_{C_d}>f_{C_o}$ difference. We have used parameter values that yield approximately the same overall level of cooperation ($f_C$) in all three cases: $\rho=0.3$, $b=1.05$ and $\alpha=0.8$ for panel (a), $\rho=0.3$, $b=1.0$ and $\alpha=0.25$ for panel (b), and $\rho=0.3$, $b=1.09$ and $\alpha=0.25$ for panel (c).}
\end{figure}

In the opposite case, when distinguished players are endowed with the full teaching activity $w=1$ while ordinary players without an external link are depreciated with $w=0.05$, the change from $\alpha=0$ to $\alpha=0.5$ is the strongest [panel (c)]. Here the inequality of $w$ and the additional payoffs stemming from the other network are able to strengthen each other and fortify the leaders to yield the highest cooperation level within the framework of this model. The enforced role of distinguished players will cause an immediate reaction from the followers, and this prompt reaction will select the more successful cooperative strategy as described before. From the technical point of view, it is interesting to note that the curves for $\alpha=0$ in Figs.~\ref{fC_comp}(b) and (c) coincide, because in the absence of extra payoff from the other network it does not matter whether distinguished or ordinary players are endowed with  $w=1$ (or $w=0.05$). Since the fraction of distinguished players was set to $\rho=0.5$, both fractions are equally large, thus resulting in the same cooperation level.

\begin{figure}[b]
\centerline{\epsfig{file=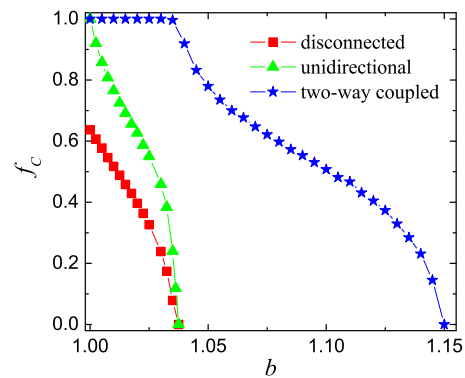,width=6cm}}
\caption{\label{one_way} Optimal interdependence between the two networks works optimally only if the links connecting them go both ways. Depicted is the fraction of cooperators $f_C$ in dependence on the temptation to defect $b$ as obtained on an isolated network (reference case), on two networks connected by means of unidirectional links, and on two mutually interdependent networks. Cooperation is optimally promoted only if there is independent formation of cooperative patterns on each individual network, for which the chance for heterogeneity (distinguished players) needs to be provided on both of them. Parameter values used were: $\rho=0.5$ and $\alpha=0.5$.}
\end{figure}

Results presented in Fig.~\ref{fC_comp} and the pertaining interpretation can be corroborated by comparing the evolution of cooperation within different groups of players, similarly as done in Fig.~\ref{evol_frac} above. As Fig.~\ref{evol_comp}(b) shows, when the teaching activity is reduced for distinguished players the cooperation among them is not favored by natural selection. Consequently, only a very low value of $b$ can ensure a reasonably high cooperation level in the whole population because the latter is dragged down considerably by the low $f_{C_d}$. There is a slight improvement among other players, as evidenced by the higher $f_{C_o}$, which is due to a higher teaching activity. However, despite their higher teaching activity, ordinary players cannot lead the whole population efficiently because distinguished players are reluctant to follow them due to their higher individual utility. If either the teaching activity of all players is left intact [panel (a)] or the higher teaching activity is awarded to distinguished players [panel (c)], then $f_{C_o}<f_{C_d}$, as observed initially in Fig.~\ref{evol_frac}. Thus, the basic mechanism of cooperation promotion is restored or even additional fortified.

\begin{figure}
\centerline{\epsfig{file=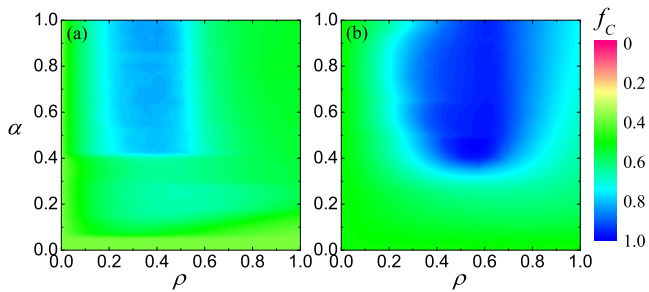,width=8.5cm}}
\caption{\label{updates} Robustness of optimal network interdependence on different strategy updating rules. Color coded is the fraction of cooperators $f_C$ in dependence on the fraction of players that are allowed to form an external link $\rho$ and the strength of these links $\alpha$, as obtained for $b=1.4$ and best-takes-over strategy updating (a), and $b=1.06$ and proportional imitation (b). Irrespective of the applied strategy updating rule, there exists an intermediate value of $\rho$ at which cooperation is optimally promoted, and the dependence on $\alpha$ is also qualitatively the same as in Fig.~\ref{ar_plane}, where the Fermi strategy adoption rule has been used.}
\end{figure}

Furthermore, it is important to emphasize that the optimal interdependence between the networks can work in favor of cooperation only if there are distinguished players in both graphs. On the other hand, if the additional utilities flow only in one direction, the mechanism will fail or yield only a marginally better outcome. Results supporting this argumentation are presented in Fig.~\ref{one_way}. For reference, the outcome on an isolated network is depicted as well. It can be observed that if distinguished players in the first network can collect additional payoffs from the second network, but the evolution in the second network is independent from the first network because players there are unable to collect additional payoffs (the interdependence is unilateral), then the critical temptation to defect $b$ at which cooperators die out does not increase at all. In fact, only the level of cooperation increases slightly in the mixed $C+D$ phase. If the interdependence is bilateral, i.e., players on both networks can collect an additional payoff from the other network, however, the positive impact on the evolution of cooperation is much stronger. This highlights the importance of mutual interdependence and the independent formation of cooperative patterns on each individual network. Players connected to their corresponding partners in the other network can support each other effectively only if homogenous cooperative domains emerge on both networks. Notably, the importance of correlated growth and formation of cooperative domains has been emphasized already in \cite{wang_z_srep13}, although here they need not overlap geographically. We could observe practically the same cooperation level when distinguished players collected additional payoffs from an ordinary player in the other network. Namely, there is no need to link distinguished players from different networks with one another. The crucial condition is the chance of heterogeneity on both networks \cite{santos_jtb12}, the positive effect of which can then be mutually amplified through the interdependence.

Our argument for the spreading of cooperative behavior among distinguished players can also be supported by how the border of the full $D$ phase behaves in dependence of $\rho$ and $\alpha$ at a high temptation to defect. As results presented in Fig.~\ref{ar_plane}(c) suggest, a smaller density of distinguished players can be compensated by a higher value of $\alpha$, but only up to a certain point. At such a high temptation value cooperators cannot survive even for large values of $\alpha$ if the density of distinguished players $\rho$ is below a threshold value. Naturally, the critical $\rho$ depends slightly on the temptation to defect [compare panels (b) and (c)], but the smallest value of the phase transition point is close to the critical $\rho_c=0.1869(1)$ value of jamming coverage of particles during a random sequential adsorption when nearest and next nearest neighbor interactions are excluded on a square lattice \cite{dickman_jcp91}. Shortly, if distinguished players are too rare, then to support them via a high $\alpha$ will not yield the desired impact because their influence cannot percolate. The latter, however, is an essential condition to maintain cooperation, which was already pointed out in previous works \cite{wang_z_pre12b, wang_z_srep12}.

\begin{figure}
\centerline{\epsfig{file=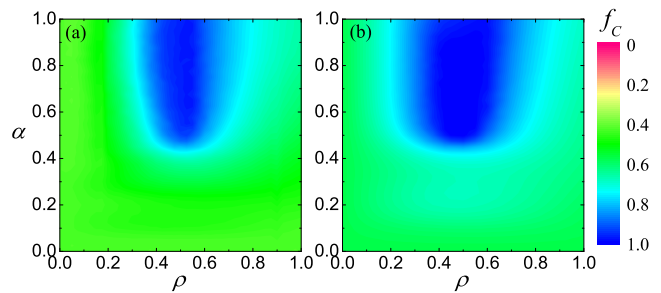,width=8.5cm}}
\caption{\label{tri} Robustness of optimal network interdependence on the topology and the type of social dilemma. Color coded is the fraction of cooperators $f_C$ in dependence on the fraction of players that are allowed to form an external link $\rho$ and the strength of these links $\alpha$, as obtained for $b=1.1$ on the triangle lattice (a), and $r=0.3$ (snowdrift game) on the square lattice (b). The Fermi strategy adoption rule has been used in both cases. Irrespective of the topology of each individual network and the type of the governing social dilemma, the results are qualitatively the same as in Figs.~\ref{ar_plane} and \ref{updates}.}
\end{figure}

It is lastly of interest to verify the robustness of these observations, first with regards to the strategy updating rule. As results presented in Fig.~\ref{updates} demonstrate, our conclusions are not restricted solely to the Fermi-type strategy updating \cite{blume_l_geb93, szabo_pr07}, but remain valid also under the best-takes-over rule \cite{nowak_n92b} and proportional imitation \cite{schlag_jet98}. In both cases an optimal intermediate value of $\rho$ is clearly inferable, and the positive effect on the evolution of cooperation is the stronger the larger the value of the coupling strength $\alpha$. This is qualitatively identical as observed in Fig.~\ref{ar_plane} with the Fermi rule.

Since previous works revealed that the clustering coefficient could be a decisive factor affecting the evolution of cooperation in games that are governed by pairwise interactions \cite{szabo_pre05, kuperman_pre12} (note that this is not the case for games governed by group interactions \cite{szolnoki_pre09c}), it is also instructive to examine the relevance of network  interdependence under this condition. Unlike the square lattice that has a zero clustering coefficient, the triangular lattice has a high clustering coefficient, and thus serves the purpose very well. Results presented in Fig.~\ref{tri}(a) attest to the fact that the existence of optimal network interdependence does not depend on structural properties of each individual network, as indeed the $\alpha-\rho$ dependence of $f_C$ is the same as observed before for the square lattice in Fig.~\ref{ar_plane}. To conclude, we extend our exploration also to other social dilemmas, more precisely the snowdrift game, and as shown in Fig.~\ref{tri}(b), the main conclusions remain intact. To reiterate, the exists and intermediate level of interdependence between networks that is optimal for the resolution of social dilemmas.

\section*{Discussion}
Summarizing, we have studied the evolution of cooperation in the spatial prisoner's dilemma and the snowdrift game subject to interconnectedness by means of a different fraction of differently strong links between the corresponding players residing on the two interdependent networks. We have found that for cooperation to be optimally promoted, the interdependence should stem only from an intermediate fraction of links connecting the two networks, and that those links should affect the utility of players significantly. The existence of optimal interdependence has been attributed to the heterogeneity that is brought about by the enhanced utility of those players that do have external links to the other network, as opposed to those who have not. This introduces asymmetric strategy flow, which in turn leads to the emergence of influential leaders that can act as strong cooperative hubs in their respective networks. Importantly, the compact cooperative patterns that appear independently on both networks support each other mutually through the links that constitute the interdependence. Indeed, we have shown that the mechanism works best only if the interdependence is bilateral, and if the asymmetric strategy flow is not counteracted by artificial low weights assigned to the reproducibility of interconnected players. In case of unidirectional interdependence or if the reproducibility of players is altered, however, some marginal benefits for cooperators may still exist, but these are then far removed from the full potential of interdependent networks to aid the resolution of social dilemmas. We have tested the robustness of these conclusions by replacing the Fermi strategy adoption rule with the best-takes-over and the proportional imitation rule, as well as by replacing the square lattice having zero clustering coefficient with the triangle lattice that has a much higher clustering coefficient, as well as finally by replacing the prisoner's dilemma game with the snowdrift game. Quite remarkably, we have found that the optimal interdependence persist across all these different setups, thus leading to the conclusion that it ought to be to a large degree a universally valid phenomenon.

While the games on interdependent networks studied here are not meant to model a particular real-life situation, they nevertheless do capture the essence of some situations that are viable in reality. For example, it is generally accepted that not all individuals are equally fond of making connections outside of their natural environment. Similarly, some would very much wish to do so, but may not have a chance. These and similar considerations may all affect the level of interdependence between two or more networks, and it is within the realm of these possibilities that our study predicts the existence of an optimal level of interconnectedness. Future studies could address the coupling of more complex (small world or scale-free, for example) interaction topologies, the outcome of other games on interdependent networks, such as for example the traveler's dilemma game that has recently been studied in a spatial setting \cite{li_rh_pone13}, as well as the impact of coevolution and growth, both of which have recently been the subject of much interest \cite{poncela_ploso08, poncela_njp09, portillo_pre12, li_g_pa12, brede_pone13}. Overall, it seems safe to conclude that the interdependence of interaction networks offers several exciting possibilities for further research related to evolutionary games, and it ought to bring the models a step closer to actual conditions, given that networks indeed rarely exist in an isolated state.

\section*{Methods}
The evolutionary game is staged on two square lattices, each of size $L \times L$, where initially each player $x$ is designated either as a cooperator $(s_x=C)$ or defector $(s_x=D)$ with equal probability. Likewise randomly, a fraction $\rho$ of players on each lattice is selected and allowed to form an external link with a corresponding player in the other lattice. Although we predominantly use the square lattice, we will also resort to using the triangle lattice, given that the difference in the clustering coefficient has been determined as a potentially key factor for the outcome of games that are governed by pairwise interactions \cite{szabo_pre05, kuperman_pre12}.

The accumulation of payoffs $\pi_x$ on both networks follows the same standard procedure, depending on the type of the governing social dilemma. In both games two cooperators facing one another acquire $R$, two defectors get $P$, whereas a cooperator receives $S$ if facing a defector who then gains $T$. The prisoner's dilemma game is characterized by the temptation to defect $T=b$, reward for mutual cooperation $R = 1$, and punishment $P$ as well as the sucker's payoff $S$ equaling $0$, whereby $1 < b \leq 2$ ensures a proper payoff ranking \cite{nowak_n92b}. We note that qualitatively similar results are obtained also for other values of $S$. The snowdrift game, on the other hand, has $T=\beta$, $R=\beta-1/2$, $S=\beta-1$ and $P=0$, where the temptation to defect can be expressed in terms of the cost-to-benefit ratio $r=1/(2\beta -1)$ with $0 \leq r \leq 1$. Due to the interdependence (external links between corresponding players), however, the utilities used to determine fitness are not simply payoffs obtained from the interactions with the nearest neighbors on each individual network, but rather $U_x = \pi_x + \alpha \pi_{x^\prime}$ for players that have an external link, and $U_x = \pi_x$ otherwise. The parameter $0 \leq \alpha \leq 1$ determines the strength of external links, i.e., the larger its value the higher the potential increase of utility of two players that are connected by the external link.

Importantly, while the interdependence affects the utility of players, it does not allow strategies to be transferred between the two networks. Thus, the evolution of the two strategies proceeds in accordance with the standard Monte Carlo simulation procedure comprising the following elementary steps for each network. First, a randomly selected player $x$ acquires its utility $U_x$ by playing the game with all its nearest neighbors and taking into account also the potential addition to the utility stemming from the possible external link, as described above. Next, one randomly chosen neighbor of $x$ within the same network, denoted by $y$, also acquires its utility $U_y$ in the same way. Lastly, player $x$ attempts to adopt the strategy $s_{y}$ from player $y$ with a probability determined by the Fermi function
\begin{equation}
W(s_{y} \rightarrow s_{x})=\frac{1}{1+\exp[(U_x-U_y)/K]} \,,
\label{fermi}
\end{equation}
where $K=0.1$ quantifies the uncertainty related to the strategy adoption process \cite{blume_l_geb93, szabo_pr07}. The latter is usually associated with errors in decision making and imperfect information transfer between the players. Notably, to test the robustness of our findings, we will also use the best-takes-over \cite{nowak_n92b} and the proportional imitation \cite{schlag_jet98} strategy updating rule. Regardless of which type of interaction network, evolutionary game, or the strategy updating rule is used, in accordance with the random asynchronous update, each player on both networks is selected once on average during a full Monte Carlo step. Moreover, sufficiently large system sizes (from $L=200$ to $L=800$) and relaxation times need to be used to avoid finite size effects and to ensure a stationary state has been reached. Presented results were averaged over up to $30$ independent runs to further improve accuracy.

\begin{acknowledgments}
This research was supported by the Hungarian National Research Fund (Grant K-101490), TAMOP-4.2.2.A-11/1/KONV-2012-0051, and the Slovenian Research Agency (Grant J1-4055).
\end{acknowledgments}

\end{document}